\documentclass[times, 10pt,twocolumn]{article} 

\usepackage{latex8}
\usepackage{times}

\usepackage{xspace}
\usepackage{graphicx}
\usepackage{subfigure}
\usepackage{paralist}
\usepackage{relsize}
\usepackage{amsmath, amsthm, amssymb}

\newcommand{\figref}[1]{Fi\-gure~\ref{#1}}
\newcommand{\tabref}[1]{Ta\-ble~\ref{#1}}

\newcommand{\sectref}[1]{Sec\-tion~\ref{#1}}

\newcommand{\tool}[1]{\textsc{#1}}
\newcommand{\commercialtool}[1]{\textsc{#1}}

\newcommand{\xunit}[0]{\tool{xUnit}\xspace} 
 
\newcommand{\cppfamix}[0]{\commercialtool{cpp2famix}\xspace}
\newcommand{\aant}[0]{\tool{Apache Ant}\xspace}
\newcommand{\ant}[0]{\tool{Ant}\xspace}
\newcommand{\junit}[0]{\commercialtool{JUnit}\xspace}

\newcommand{\SSVT}[0]{S-S-V-T\xspace}

\newcommand{\vis}[0]{visualization\xspace}

\newcommand{\testhelper}[0]{\emph{test helper\xspace}}
\newcommand{\testhelpers}[0]{\emph{test helpers\xspace}}
\newcommand{\swview}[0]{System-Wide View}

\newcommand{\question}[2]{}

\newcommand{\material}[1]{}
\newcommand{\trial}[1]{}
\newcommand{\removed}[2]{}
\newcommand{\musings}[1]{}
\newcommand{\sketch}[1]{}
\newcommand{\comment}[1]{}
\newcommand{\vocabulary}[2]{}
\newcommand{\latextask}[2]{}

\newcommand{\relatedwork}[1]{}

\newcommand{\cvsversion}[1]{}

\newcommand{\head}[1]{\vspace{+1ex}\par\noindent\textbf{#1}\ \ }
\newcommand{\oat}[0]{o.a.t}
\newcommand{\fetch}[0]{\tool{Fetch}}
\newcommand{\guess}[0]{\tool{Guess}}
\newcommand{\emma}[0]{\tool{Emma}}

\begin{document}

\title{\vspace*{-3ex}Exploring the Composition of Unit Test Suites}

\author{Bart Van Rompaey and Serge Demeyer\\
Lab On Re-Engineering\\
University Of Antwerp \\
\{bart.vanrompaey2,serge.demeyer\}@ua.ac.be
}

\maketitle
\thispagestyle{empty}


\begin{abstract}\vspace{-1ex}
In agile software development, test code can considerably contribute to the overall source code size. Being a valuable asset both in terms of verification and documentation, the composition of a test suite needs to be well understood in order to identify opportunities as well as weaknesses for further evolution. In this paper, we argue that the visualization of structural characteristics is a viable means to support the exploration of test suites. Thanks to general agreement on a limited set of key test design principles, such visualizations are relatively easy to interpret. In particular, we present visualizations that support testers in
(i) locating test cases; (ii) examining the relation between test code
and production code; and (iii) studying the composition of and dependencies
within test cases. By means of two case studies, we demonstrate how visual patterns help to identify key test suite characteristics. This approach forms the first step in assisting a developer to build up understanding about test suites beyond code reading.
\end{abstract}


\section{Introduction }
\label{sec:introduction}

Pushed by the adoption of agile development methodologies as well as the availability of free testing frameworks, a lot of unit tests have been written over the last few years. Such tests are specified persistently, thereby contributing to the size of a software project's artifacts. Studies report a ratio of test to production code which can extend till 2:3; occasionally even 1:1 \cite{DMBK2001RTC,Laplante2006TDDInLarge}. As such, unit testing considerably impacts a software project's development cost.

The benefits of unit testing are well known. In the short term, the application of unit testing results in software of higher quality \cite{Murphy94classtesting,Max03TDDatIBM,Crispin06TDDandSQ}. Unit testing is observed to find other defects \cite{Runeson03UTvsCI,ellims04testingInPractice} and is also reported to be considerably cheaper than strategies relying solely on testing later in the development cycle \cite{ellims04testingInPractice}. In the long term, unit tests are a valuable asset during regression testing, able to notice undesired side effects of changes \cite[Ch.6]{Demeyer2002Object}.

On the down side, unit test code needs to co-evolve with production code in order to remain useful. Moreover, a test suite is subject to the problem of design erosion as well, gradually loosing the initially intended design and thereby becoming harder to understand and modify. Constructs in the tests that hinder modification, e.g. complex test cases or a resource dependent test \cite{DMBK2001RTC}, directly affect developer productivity thereby amplifying the overall maintenance cost. Studies indicate that regression testing can account for as much as one-third of the total cost of a software system \cite{Harrold00TestingRoadmap}.

Therefore, in the context of a legacy system, the associated test suite contains both opportunities, in the form of well designed, isolated unit tests with a high coverage, as well as weaknesses, in the form of maintenance intensive test cases, components lacking coverage, etc. Evaluating the overall condition first requires identifying the location of test code and relating it to the corresponding production code. A first notion of coverage per component can be obtained by comparing production and test code size-wise. Next, to explore the amount and kind of test cases for individual production components, a developer has to study their interdependencies. Detecting maintenance intensive test cases, finally, requires studying their internals: typically through code reviewing.

Code reviewing, with known reviewing rates around 150 to 200 lines of code per hour \cite{Belli1997}, was quickly found not to be scalable and therefore research went to look for design recovery techniques at a higher level of abstraction \cite{Biggerstaff89DesignRecovery,Lanza03Polym}. We identified two ways in which general design recovery techniques can exploit the more constrained context of test code. First, in contrast to the heterogeneous design heuristics for production code, design guidelines for test code are quite strict, emphasizing recurrent design idioms such as the setup-stimulate-verify-tear down cycle (\SSVT). Secondly, the abstractions typically used in program comprehension - e.g. classes, methods, invocations etc. - lack testing semantics. Testers reason in terms of test cases, fixtures and assertions, suggesting a semantic layer on top of the abstractions employed in general design recovery.

Accordingly, this work contributes to the general body of knowledge
on software visualization by introducing a test suite representation
that does exploit the more constrained context of test code, allowing developers to explore the composition of a test suite, navigating between corresponding production units and test cases, identifying co-evolution needs or spotting test design anti-patterns.

This paper is structured as follows. In \sectref{sec:testsuitestructure}, we recapitulate desired unit test characteristics. We clarify how the use of the \SSVT cycle as well as unit testing frameworks assist in composing well-structured tests. The visualization technique introduced in \sectref{sec:visualizingtestsuites} exploits software design elements in its abstraction. Next, we present three visual presentations and discuss their interpretation (\sectref{sec:threetestsuiteviews}). In \sectref{sec:casestudies}, we report about two case studies, the findings of which we validate by means of design documentation, reports as well as an interview with a developer. After discussing related work (\sectref{sec:relatedWork}) we wrap up (\sectref{sec:conclusion}).

\section{Test Suite Design}
\label{sec:testsuitestructure}
In this section, we briefly introduce terminology, design guidelines and strategies that are commonly used during unit testing.

\paragraph{Unit Testing Terminology --}The standard unit testing terminology stems from Beck's pattern system \cite{BeckTestPatterns}:
\begin{compactitem}
	\item a \emph{Unit under Test} is the set of production classes (classes that contribute to the final software product) that is exercised together during testing. In a strict unit testing approach a unit corresponds to a single class.
	\item a \emph{Test Case} groups a set of tests performed on the same unit under test. Within \junit\footnote{\junit is the Java implementation of the \xunit family of testing frameworks, the \emph{de facto} framework for unit testing today}, a test case is specified as a class, inheriting from the generic \texttt{TestCase} class offered by the test framework.
	\item a \emph{Test Case Fixture} is the set of attributes a test case requires to bring the unit under test into the desired state. The fixture consists of an instance of the unit under test as well as some shared test data. 
	\item a \emph{Test Command} is a container for a single test. It is typically encapsulated in a method of a test case.
	\item the \emph{Test Case Setup} is a method of the test case in which the fixture is initialized into the desired state for testing. A corresponding \emph{Test Case Tear Down} method releases resources again.
\end{compactitem}

\paragraph{Design Guidelines --} Unit test design guidelines propose a strict structure for specifying tests: (i) acquire and initialize the necessary resources, (ii) send one or more \emph{stimuli} to the unit under test, (iii) verify that the unit responds properly; and finally (iv) release the acquired resources. These four calls are referred to as the setup-stimulate-verify-tear down cycle (\SSVT). The first step, performed in a Test Case Setup, is repeated before every Test Command in the Test Case. Each Test Command stimulates and verifies the unit under test.

Unit tests are typically specified in the same programming language as the system under test, yet are not tested extensively themselves. To support code reviewing as the main means for test quality assurance, test cases are required to be \emph{concise}, \emph{transparent in their objectives} and \emph{isolated} in implementing the \SSVT cycle, forming an \emph{encapsulated} test. Test Commands exercising the same unit under test are gathered in a Test Case, thereby sharing an \emph{explicit} Test Case Fixture and Setup. 

\paragraph{Unit Testing Strategies --} The testing plan of a typical software system entails many strategies: unit tests verify the functionality of small units at a time, integration tests are focused on the interaction between components, system tests consider the behaviour of the system overall, etc. Despite the clearly distinctive objectives for each strategy, overlap between strategies occurs, such as unit tests bearing properties of another testing strategy:
\begin{compactitem}
	\item Units that are tightly coupled with many other units require more effort to isolate, e.g. by means of test stubs taking the place of external units. Therefore, a tester may decide to unit test without isolating - i.e. setting up a larger unit under test only part of which is the actual unit under test. Such a test case can be considered as being more integration testing.
	\item Certain units are always used by the same other (set of) units. During testing that unit might also be exercised via this set for ease of setup or because it closer resembles the actual usage scenario, resulting in multiple units to be considered. This testing approach is called Indirect testing. Moonen and van Deursen argue that this makes understanding and debugging harder \cite{DMBK2001RTC}.
	\item When large data sets are required for testing certain units, this data is sometimes stored in files and loaded during test setup. The reading functionality is typically abstracted and shared among test cases. Test code classes that implement this functionality are referred to as \emph{test helpers}.
	\item System-wide input/output testing approach can be fed with test input data specifically chosen to exercise particular units.
\end{compactitem}
Considering this variation, determining the kind of unit tests present in a test suite is worthwile, as it may steer upcoming re-engineering tasks.

\section{Visualizing Test Suites}
\label{sec:visualizingtestsuites}

In this work, we propose a visualization technique assisting re-engineers to analyze and comprehend the structure and quality of the test suite of large systems. To describe our visualization, we use the five-dimensional framework of Maletic et al. \cite{833803}. This framework was proposed for development and maintenance of large-scale software and stimulates the user to describe how a technique assists in completing a particular software development task. 

\subsection{Tasks}

The proposed visualization supports three program understanding tasks that we call \emph{First Contact}, \emph{Understand Unit(s)} and \emph{Assess Test Case(s)}. Together, they form a top-down and phased approach to explore a test suite.

\head{Perform First Contact.} In this task, a developer builds up an overall mental model of a system at a high level of abstraction \cite{Demeyer2002Object}. Initial understanding of the associated test suite is obtained by: (i) localizing the test suite code in the source tree, (ii) looking at overall coverage to get a notion on covered\footnote{In the context of this paper, we use a coarse grained definition for coverage: a class is covered by a test case when at least one of its methods is invoked by a test command} as well as uncovered components; and (iii) studying test suite design to grasp the granularity of units that have been used.

\head{Understand Unit(s).} Koenemann and Koch observed that programmers only study code in case they are convinced of the relevance in the context of a particular task \cite{Koenemann91}. As unit tests also serve the purpose of live documentation, explaining in simple scenario's how a unit is (and is not) supposed to be used, information about relevant test cases forms a next step towards the actual modification of the code. Secondly, coverage information can reveal important parts, as we assume that critical, frequently changing or core parts are tested more extensively.

\head{Assess Test Case(s).} After having identified relevant unit tests, i.e. test cases directly invoking production units of interest, the next step consists of evaluating the internal structure of individual test cases. Well-designed unit tests are most suited for documentation purposes, as they (i) are easy to understand; and (ii) specify how a particular unit is (and is not supposed to be) used in an isolated scenario. Test cases may present certain maintenance challenges as well. Weakly isolated test cases, requiring a complex setup or bearing unrealistic run-time expectations, become costly to maintain due to frequent changes, slow execution or seemingly random failures \cite{DMBK2001RTC}.

\head{}Summarizing, for three maintenance tasks we identified three information topics concerning the test suite: Test Location, Test Coverage and Test Design.\tabref{tab:topicview} shows which information topics are required during the three tasks. For each of them we will introduce a separate view, i.e., a filter mechanism on the overall visualization technique.

\begin{table}[htb]
\centering
	\begin{smaller}
	\begin{tabular}{ l | c | c | c }
	\hspace{35pt}\emph{Task} & \emph{Perform} & \emph{Understand} & \emph{Assess} \\
	\emph{Topic} & \emph{First Contact} & \emph{Unit(s)} & \emph{Test Case(s)} \\
	\hline
	Test Location & $\surd$ & $\surd$ & \\
	Test Coverage & $\surd$ & $\surd$ & \\
	Test Design & $\surd$ & & $\surd$ \\
	\end{tabular}
	\end{smaller}
\caption{How are test suite exploration topics covered by the presented views?}
\label{tab:topicview}
\end{table}

\subsection{Audience}

The visualization technique is intended to assist software engineers in exploring the test suite of an unfamiliar system, as required in the following typical scenarios: 
\begin{compactitem}
	\item A newcomer to the project who is asked to build up knowledge quickly and relatively independent from the existing team. 
	\item A team of developers assigned to a major modification of a stabilized, long running legacy system may use it to characterize the available tests, thereby serving as a reference for communication.
	\item A re-engineer, asked to analyze a system X's opportunities and threats when considered to be integrated into system Y, will be interested in the test suite as well.
\end{compactitem}

\subsection{Target}

The target defines the characteristics of the software system to be visualized \cite{833803}. In this work, we are interested in the structure of test suites as well as the relationship between test suites and production code. In a first step, we fetch information from the system's source code using a static fact extractor. This results in a model of the system according to the formalism of the Object Oriented Framework for Coupling and Cohesion (OOFCC) specified by Briand et al. \cite{Briand1999}, i.e. a formalism to query and count in terms of classes, methods, invocations, etc. Secondly, we identify entities that belong to test code and adapt the model representation according to the OOFCC refinement that we proposed for test code that (i) formalizes unit test concepts as entities and relationships in a test model, (ii) describes how OOFCC entities map onto test concepts for common implementations of xUnit and (iii) provides the heuristics (type, inheritance and ownership properties as well as naming conventions) required to implement a refining model transformation \cite{VanRompaey2006TestSmellMetrics}. Due to space constrains, we do not reproduce the formalism here. Next, for each view we query this model and compose the input for a graph visualization tool. These steps are automated in a tool called \fetch\footnote{stands for Fact Extraction Tool CHain, developed at the University of Antwerp. Available at http://www.lore.ua.ac.be/Research/Artefacts/}.

The model entities of \cite{VanRompaey2006TestSmellMetrics} were already introduced (informally) in Section~\ref{sec:testsuitestructure} (Unit Testing Terminology). We furthermore distinguish three types of relations:
\begin{compactitem}
	\item \emph{Containment} relations represent the hierarchical decomposition of a software system in containers. Classes belong to a parent package, methods belong to a class, etc. This decomposition applies to both production and test code.
	\item A \emph{coverage} relation is a relation between a test entity \emph{A} and a production entity \emph{B}, where \emph{A} has at least one invocation towards \emph{B}. As such, we only consider direct relations between production and test code, in line with a developer's information needs during the exploration of a test suite's composition, rather than obtaining finer-grained, exact coverage measures.
	\item \emph{Test dependencies} are relations between test entities, e.g., revealing certain abstractions or recurrent helper functions in the test design. In practice, such mechanisms are typically implemented in an (abstract) test case superclass. As such, we denote an inheritance relation between two test cases as a test dependency.
\end{compactitem}

\subsection{Representation \& Medium}

Directed graphs have been shown to be natural representations of software systems \cite{Gulla1992MVVis,ball97Vsystem}. \tabref{tab:legend} explains which entities and relations we represent as nodes and edges. The symbols for class and test case vary per view. For the 'medium' dimension, characterizing where the visualization is to be rendered \cite{833803}, we adopted \guess. \guess\emph{ } is an \emph{exploratory data analysis and visualization tool for graphs and networks} \cite{1124889}, as a part of \fetch. This environment assists the user in graph exploration through capabilities such as applying graph layouts, highlighting, zooming and moving as well as customizable filtering. 

\begin{table}[htb]
	\centering
	\begin{smaller}
	\begin{tabular}{| l | c | c | c |}
	\hline
	\emph{Model Entity} & \emph{Type} & \emph{Prod/Test} & \emph{Repr.} \\
	\hline
	Package 	& Node 	& Prod  &   \includegraphics[scale=0.2]{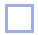} \\
	Class		& Node   	& Prod 	&   \includegraphics[scale=0.2]{images/packageNode}/\includegraphics[scale=0.2]{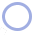} \\
	Method		& Node  	& Prod  &   \includegraphics[scale=0.2]{images/methodNode} \\
	Test Case 	& Node 	& Test  &   \includegraphics[scale=0.35]{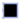}/\includegraphics[scale=0.2]{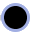} \\
	Test Command 	& Node	& Test	&   \includegraphics[scale=0.2]{images/testCommandNode} \\
	Containment 	& Edge   	& -     &   \includegraphics[scale=0.2]{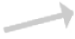} \\
	Coverage 	& Edge   	& -     &   \includegraphics[scale=0.2]{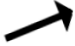} \\
	Dependency 	& Edge   	& -     &   \includegraphics[scale=0.2]{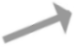} \\
	\hline
	\end{tabular}
	\end{smaller}
\caption{Visualization Legend}
\label{tab:legend}
\end{table}

\section{Three Test Suite Views}
\label{sec:threetestsuiteviews}

In this section we present three views as filters on the overall graph representation introduced above. Each view corresponds to an exploration task. We expand upon the intent and motivation for each view, and the interpretation that should be given to any of the observed indicators.

\subsection{System-wide Test Suite View}

The core of this \vis is the hierarchical decomposition of a software system into packages and classes. The filter we apply on the graph therefore skips all entities below the class level. To distinguish packages entities from class entities we use square and circle shapes respectively. The Graph EMbedder (GEM) algorithm applied on the containment edges provides both an easy to interpret as well as aesthetically pleasing layout \cite{Frick94GEM}. 

In line with the re-engineering pattern "Study the exceptional entities" \cite{Demeyer2002Object}, entities exhibiting either a lack of, oppositely, plenty of incoming and outgoing edges deserve special attention. In the former case, one has to look further for other testing strategies. In the latter case, an important role during testing can be assumed.

\subsubsection{Test Location}

\head{Intent.} Localize the system test code.

\head{Motivation.} Typically, localization of test code is a configuration management responsibility, as it is a consequence of source tree structure in the version control system. Feathers discusses pro and contras of possible test code localization: the whole test suite may be gathered in a common location, test cases may be stored per component, but may as well reside among production code \cite{Feathers2005}.

\head{Interpretation.} A consistent location means that we identified earlier project conventions that we can rely on. In case of no dominant visual indicators, we can deduce the absence of such conventions. The view still helps to determine the test code associated with a particular component.   

\begin{figure}[t]%
	\centering
	\subfigure[Same Package]{
		\label{fig:swDetaila}
		\includegraphics[scale=0.2]{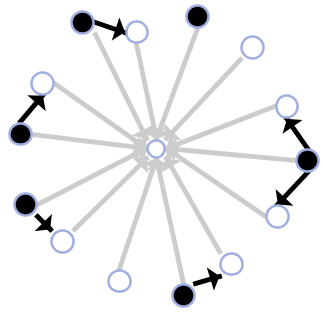}
	}	
	\subfigure[Different Package]{
		\label{fig:swDetailc}
		\includegraphics[scale=0.2]{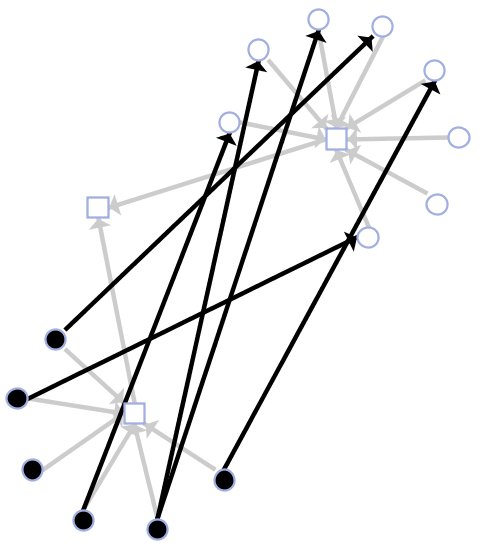}
	}
	\vspace{-4mm}
	\caption{Possible locations of test cases}%
	\label{fig:swDetail}
	\vspace{-4mm}
\end{figure}

\head{Visual Indicators.} We demonstrate two kinds of locations deducible from the system-wide view (see \figref{fig:swDetail}). 

\begin{compactitem}
	\item Test cases located in the \emph{same package} as production code will result in package nodes containing both white and black nodes. 
	\item We observe two packages, one filled with white nodes next to one with black nodes, connected by coverage edges when test code resides in \emph{another package} than production code. 
\end{compactitem}

\subsubsection{Test Coverage}

\head{Intent.} Obtain a basic notion of test coverage. 

\head{Motivation.} The desired notion of test coverage is cheap to obtain and scalable. It helps to make assumptions about earlier test efforts and system-critical components. Observing coverage for evolving components gives an impression about the risks (and possible counteractions) for further modifications. 

\head{Interpretation.} Components that are not directly tested might be trivial (e.g. data holders), have been decided only to be tested in conjunction with other components or might not have been tested, e.g., due to time constraints. For stronger tested components we assume a more important role such as being critical, frequently changing, belonging to the system core, etc. Assumptions need to be verified further on in subsequent exploration tasks, eventually by obtaining actual, fine-grained coverage measures.

\begin{figure}[htb]%
	\centering
	\subfigure[Untested Components]{
		\label{fig:swTCovNo}
		\includegraphics[scale=0.15]{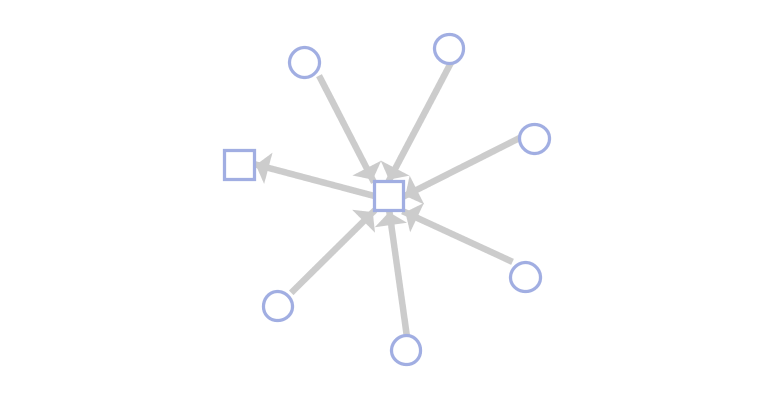}
	}	
	\subfigure[Highly Covered Class]{
		\label{fig:swTCovKey}
		\includegraphics[scale=0.2]{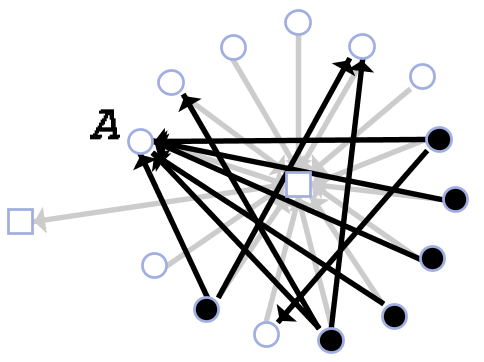}
	}
	\vspace{-4mm}
	\caption{Test Coverage Indicators}%
	\label{fig:swTestCoverage}
	\vspace{-4mm}
\end{figure}

\head{Visual Indicators.} \emph{Components not covered} by unit tests will show as clusters of classes (i) without test cases in the same package; and (ii) without incoming test coverage edges (e.g., \figref{fig:swTCovNo}). The components in \figref{fig:swDetail} serve as examples of better covered components. \emph{Highly covered classes}, such as class \emph{A} in \figref{fig:swTCovKey}, are represented as nodes receiving many test coverage edges.

\subsubsection{Test Design}

\head{Intent.} Grasp the overall test suite design.

\head{Motivation.} The test design reveals first hand information on what kind of testing strategies have been applied in the past. This tells us at which point such tests become most useful as well as how difficult tests will be to understand and modify.

\head{Interpretation.} To grasp testing strategies, we mainly look at the size of a unit and the presence of test helpers. Units of a limited size (e.g. a single class) can play an import role as test harness for local changes, due to being easier to understand and modify. When units are larger or when dealing with a more integration testing style, tests are more suited to give feedback about the overall status of a modified system. Such tests risk to be more complex and change sensitive, however, due to the many dependencies. Test helpers are typically used to abstract away recurring setup, stimulate or verification behavior, but also to facilitate access to more complex test data and input/output tests. As reusable entities, test helpers help to avoid duplication in test code.

\begin{figure}[htb]%
	\centering
	\subfigure[An isolated (A) and a less isolated component (B)]{
		\label{fig:swIsolation}
		\includegraphics[scale=0.2]{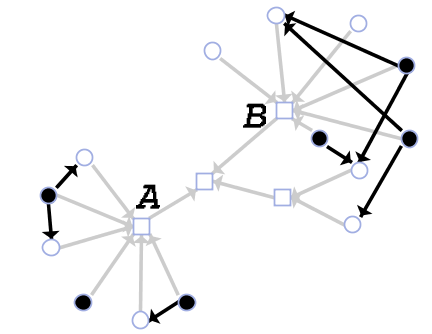}
	}	
	\subfigure[Indirect Tests via interface]{
		\label{fig:swIndirect}
		\includegraphics[scale=0.2]{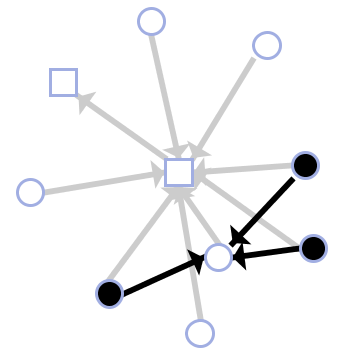}
	}
	\subfigure[Test Dependency]{
		\label{fig:swDependency}
		\includegraphics[scale=0.15]{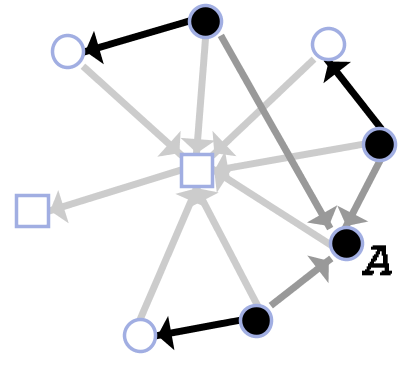}
	}
	\vspace{-4mm}
	\caption{Test Design Indicators}%
	\label{fig:swTestDesign}
	\vspace{-4mm}
\end{figure}

\head{Visual Indicators.} \emph{Units tested in isolation} are shown as package-clustered nodes receive a limited number of edges from similarly clustered test case nodes (\figref{fig:swIsolation}). We identify \emph{indirect tests} in case test cases do not cover the units expected from the identified test location, but rather  (i) access units in other packages or (ii) multiple test cases target a single unit in a package as in \figref{fig:swIndirect}.  A test case receiving many test dependency edges, such as \emph{A} in \figref{fig:swDependency}, serves, at least partially, as \emph{test helper} and forms as such an opportunity during test suite extension.

\subsection{Unit under Test View}

This view focuses on an individual production class visualized in terms of its accessible methods. Test case invoking one or more methods of this unit are displayed as well, with coverage edges drawn from the test commands to the involved methods. To distinguish classes from their methods, these entities are drawn as squares and circles respectively.

\subsubsection{Test Location}

\head{Intent.} Identify test cases for a particular unit under test.

\head{Motivation.} During evolution of component, a developer gathers the set of production classes as well as test cases that require modification.

\head{Interpretation.} Depending on past test strategies, a unit can be exercised by one or more test cases. In case unit testing as well as integration testing are specified in test code, units are covered by test cases of each strategy. Test cases might also be split up when growing too large \cite{DMBK2001RTC}. 

\head{Visual Indicators.} Trivially, production classes that are not covered directly, and as such are not likely to be the subject of a focused unit test, will show up in the view as \emph{untested units} -- components without black test nodes. Every \emph{involved test case} will be represented in terms of its test commands. Coverage edges make the relationship with the unit explicit (see \figref{fig:separatedtests}).

\subsubsection{Test Coverage}

\head{Intent.} Which parts of a unit are being tested?

\head{Motivation.} Once the scope of interest for a certain maintenance task has been reduced to a couple of units, the Unit under Test view presents how these units are covered by test commands of involved test cases. These test commands are worth exploring in detail, because of their documentation power as well as their co-evolution needs.

\head{Interpretation.} Complementary to assumptions derived from studying the location of involved test cases, in the context of coverage assessment the focus lies on identifying combinations of production methods being exercised. This allows the developer to understand which methods are not directly tested, methods covered by means of simple scenarios and methods tested together within a test command.

\begin{figure}[htbp]
	\centering
	\includegraphics[scale=0.30]{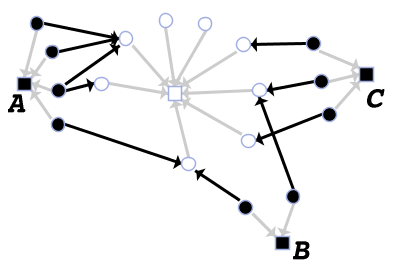}
	\vspace{-4mm}
	\caption{Multi-Test Coverage}
\label{fig:separatedtests}
	\vspace{-4mm}
\end{figure}

\head{Visual Indicators.} An example of \emph{Multi-Test Case Coverage} is shown in \figref{fig:separatedtests}: a unit receiving coverage edges from multiple test cases \emph{A}, \emph{B} and \emph{C}. If only \emph{A} would have existed, we encountered a unit with \emph{partial coverage}, i.e. only a small part of its methods are directly covered. 

\subsection{Test Case View}

The Test Case View centers around individual test cases, which are represented in terms of the \SSVT entities and the exercised production units. Again, classes are represented as squares; methods as circles. In addition to the nodes introduced in the visualization technique, this view adds two meta-nodes named \emph{Fixture} and \emph{Test Commands}, thereby making these two test concepts explicit.

\subsubsection{Test Design}

\head{Intent.} Identify opportunities in the form of well designed test cases as well as possible maintenance threats by studying the internal structure of selected test cases.

\head{Motivation.} Test cases that are designed according to strict unit test design guidelines, with explicit fixture and concise setup and test commands, are an opportunity to understand (i) typical usage of the unit under test as well as (ii) how the test suite covers such units. Integration-style test cases demonstrate how components interact. Using method-level information a developer can better motivate whether a certain test case is a possible threat or rather a manifestation of a certain test strategy.

\head{Interpretation.} Deviations from the design guidelines are potential maintenance threats. A list of complex and thus undesired test structures can be found in \cite{DMBK2001RTC}.
\begin{compactitem}
	\item Test cases can lack an explicitly defined fixture, thereby removing the distinction between the defined unit of interest and surrounding helper units.
	\item A test command invoking many production methods, possibly from multiple production classes, entails a complex (integration) test scenario. 
	\item Test cases with a large fixture that is only partially used by individual test commands indicates that the contained test commands do not logically belong together, therefore violating the guidelines of encapsulation and transparency in test objective.
\end{compactitem}

\begin{figure}[htb]%
	\centering
	\subfigure[Well-designed Test Case]{
		\label{fig:WellDesignedTest}
		\includegraphics[scale=0.2]{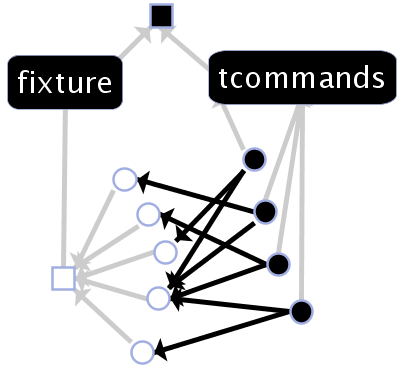}
	}	
	\subfigure[Test Case with lack of explicit fixture]{
		\label{fig:LackOfFixture}
		\includegraphics[scale=0.15]{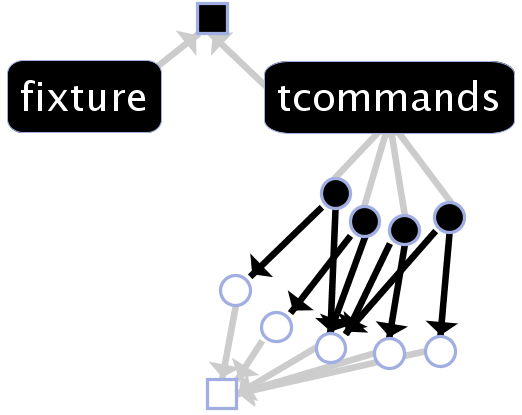}
	}
	\subfigure[Large Fixture and Complex Test Scenario]{
		\label{fig:GeneralFixture}
		\includegraphics[scale=0.22]{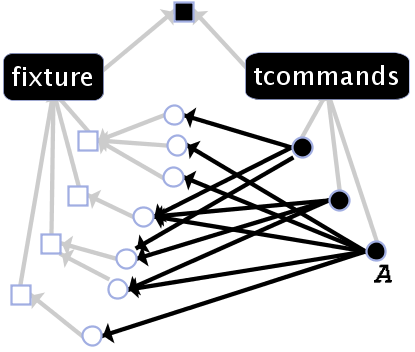}
	}
	\subfigure[Integration Test]{
		\label{fig:IntegrationTest}
		\includegraphics[scale=0.34]{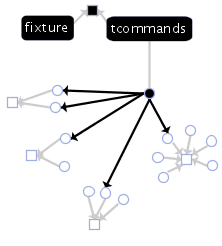}
	}
	\vspace{-4mm}
	\caption{Test Design Indicators}%
	\label{fig:TCDesignIndicators}
	\vspace{-4mm}
\end{figure}

\head{Visual Indicators.} \figref{fig:WellDesignedTest} shows an example of a \emph{well designed test case}. It contains an isolated and explicit fixture, the methods of which are consistently tested by single test commands. The \emph{Lack of Explicit Fixture} (\figref{fig:LackOfFixture}) renders a test case more difficult to understand, as the common unit under test (if at all present) is implicitly interwoven within every single test command. Making the fixture explicit implies introducing a test case attribute that is initialized by the Test Case Setup. A \emph{Complex Test Scenario} in a test command can be recognized by the many coverage edges that target production methods (e.g. test command \emph{A} in \figref{fig:GeneralFixture}). Moreover, the \emph{Large Fixture} of this test case is diagnosed by the many production class entities in the fixture that are extensively, yet not fully, shared among the test commands. Within a unit test suite, we identify the more \emph{Integration Test} type of test case (e.g., \figref{fig:IntegrationTest}) by the multiple production classes that are accessed without a dominant unit under test.

\section{Case Studies}
\label{sec:casestudies}
In this section we report on two case studies to evaluate our \vis technique. We used \fetch \emph{ }to statically extract the required information and compose the graphs. As a first project, we selected the open source build system \aant, a middle-sized, industry-strength software system. Secondly we analyzed a small system, \cppfamix, that is developed using a strict unit testing approach. This allows us to confront test suites resulting of two testing strategies. For each case study, we undertake the three identified program understanding tasks and formulate our findings based upon what we derive from the three views. Next, we validate these findings by skimming through the documentation, by looking at external references or, in the case of \cppfamix, by interviewing the developer.

\subsection{Apache Ant}

As a first case study we use the well-known \aant project. The 1.6.5 release consists of about 104 kSLOC, 18 kSLOC (17\%) of which is \junit test code.

\subsubsection{Findings}

\begin{figure*}[!t]
	\centering
	\includegraphics[scale=0.2]{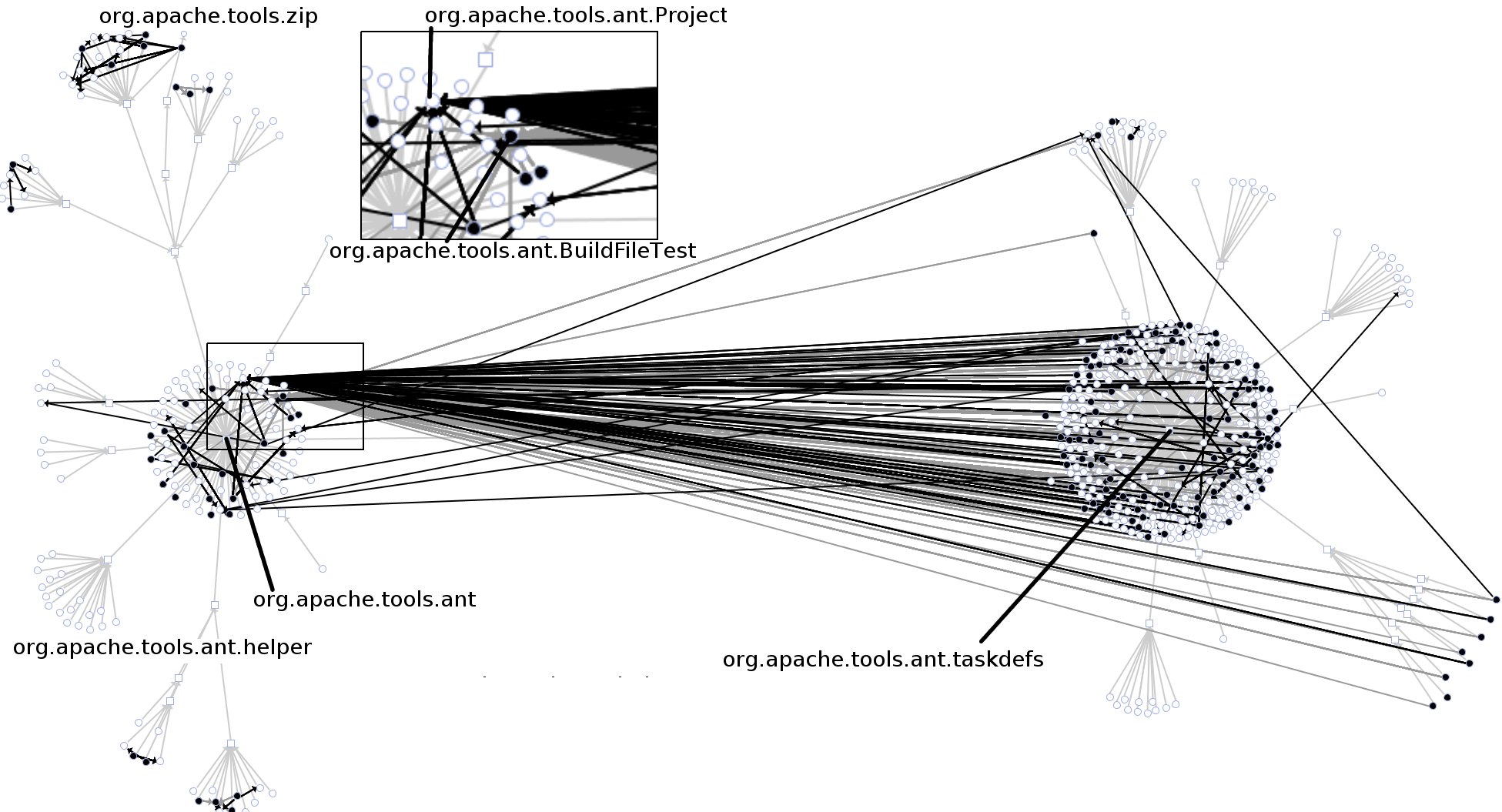}
	\vspace{-6mm}
	\caption{Ant System-wide Test View}
\label{fig:systemwide}
	\vspace{-4mm}
\end{figure*}

\head{First Contact.} Figure \ref{fig:systemwide} gives an overview of the system-wide view for test suites, based on part of the system core (due to scalability constraints on paper, we filtered out entities and relations other than the core packages \emph{\oat\footnote{abbreviation for org.apache.tools}.ant} and \emph{\oat.ant.taskdefs}). We observe the presence of a considerable amount of test cases, residing in the same packages as production code. For packages outside of the core, the testing strategy seems different: some components seem not tested at all (e.g. \emph{\oat.ant.helper}), others are tested in isolation (e.g. \emph{\oat.zip}). Although located among production classes, we notice that \ant's test cases do not cover the units in the same package. Combined with the fact that many tests are covering \emph{\oat.ant.Project} and depend upon \emph{\oat.ant.tools.BuildFileTest}, we assume an indirect testing approach. Next to \emph{\oat.ant.Project}, we identify \emph{\oat.ant.util.\{FileUtils,JavaEnvUtils\}} and \emph{\oat.ant.types.\{AbstractFileSet,Path\}} as key classes through their extensive test coverage.

\begin{figure}[!htb]
	\centering
	\includegraphics[scale=0.3]{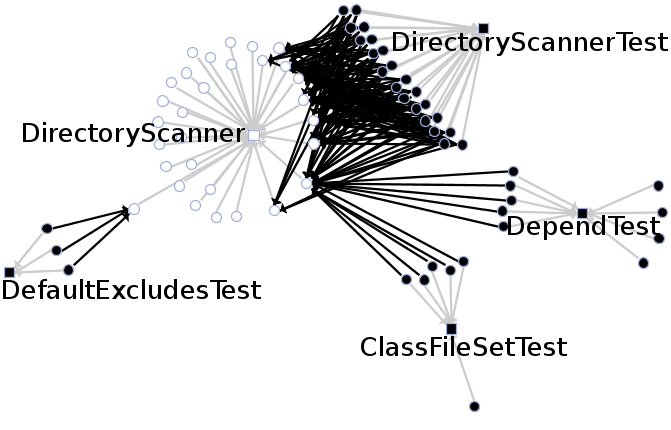}
	\vspace{-6mm}
	\caption{Unit \oat.DirectoryScanner.}
\label{fig:AntDirectoryScanner}
	\vspace{-4mm}
\end{figure}

\head{Understand Units.} In the 1.6.4 release of \ant, several bugs where found in the \emph{directory scanner} as well as the \emph{unzip} and \emph{untar} features\footnote{mentioned in the release notes of version 1.6.5}. Therefore, we analyze the existing testing facilities for these units. The concept of a directory scanner is implemented in the class \emph{\oat.ant.DirectoryScanner}. Using the Unit under Test view in \figref{fig:AntDirectoryScanner}, we note that four test cases exercise this production class: one test case is exercising eight out of twenty-two production methods, the other three invoke just two methods. Therefore, we derive that the actual unit test for this unit is \emph{\oat.ant.DirectoryScannerTest}, while the three other test cases are only using the directory scanner as a helper unit. Indeed, these test cases only invoke so called \emph{getter} methods, fetching data from the DirectoryScanner object to evaluate the expected result for their unit under test against the actual test result.

The untar functionality is implemented in \emph{\oat.ant.taskdefs.Untar}, which shows as untested. However, we do know, from the \swview, that for the \ant project test cases are located in the same package as the production classes. We identify \emph{UntarTest} as the actual test case based on naming. 

\head{Assess Test Cases.} {\oat.ant.DirectoryScannerTest} is observed to be a test command-rich test case. Most of these test commands appear to be similar in composition, targeting the same side objects of \emph{Project} and \emph{FileUtils}. \emph{UntarTest} is characterized by indirect testing behaviour by test cases relaying via \emph{BuildFileTest}, \emph{Project} and \emph{FileUtils}. Neither of the two test cases has an explicitly defined fixture. Summarizing, both test cases make use of key system classes to exercise the unit under test.

\subsubsection{Validation}

Using the code coverage tool \emma\footnote{http://emma.sourceforge.net/}, we compute a method coverage of 65\% (80\% class coverage) for \ant, confirming our initial impression of a reasonably tested system. From \ant's documentation\footnote{http://www.codefeed.com/tutorial/ant\_config.html}, we derive that the key classes in the design as identified by its architects are a.o. \emph{Project}, \emph{Task} and \emph{Target}. The documentation header of \emph{\oat.ant.Project} describes this class as the central representation of an \ant project. As this class also provides the means to start a build, its frequent usage by test cases confirms the indirect testing approach in which several project scenarios are constructed, executed and verified using this generic Project class. The focus of \ant on Java software and the fact that build instructions are specified in XML files explains the importance of the other classes we noted. By looking into one of the test cases making use of the \emph{FileUtils} class, we were able to find the XML files containing test data in the distribution. Thus, \ant's documentation confirms our assumption that production classes invoked by many test cases play a major role in the system itself. 

Our claim that \emph{\oat.ant.tools.BuildFileTest} is an important test class gets backed up by Van Geet and Zaidman, who identify it as \emph{an abstraction of a unit test that uses a build file as test data} \cite{VanGeetPCODA2006}. For each test run, a Project instance is created which loads this XML file and executes the contained build instructions. This class thus indeed serves as a \testhelper, used by no less than 414 test commands. 

The \ant case reveals a major limitation of our visualization technique: when dealing with indirect tests, we fail to trace coverage relationships between test cases and corresponding unit under test. We argue however that in a first contact phase, where overall system comprehension as well as identification of components worth investigating further are the prime objectives, information such as actual, complete coverage measurements requiring dynamic analysis are too expensive and less suited for visualization. Given the underlying test model, graph querying is also a viable alternative to reveal indirect testing relations.

\subsection{CPP2FAMIX}

As a second case study we opt for the \cppfamix, a C++ fact extractor of about 13.5 kSLOC Java code. \cppfamix extracts information about a C++ software system out of the AST unit dumps of the GNU Compiler Collection. This information is transformed into a re-engineering model. The \junit test suite accounts for 29\% of the overall system size. We chose this system because the developer is a colleague of ours, hence we can thoroughly interview him. 

\subsubsection{Findings}

\begin{figure*}[htbp]
	\centering
	\includegraphics[scale=0.40]{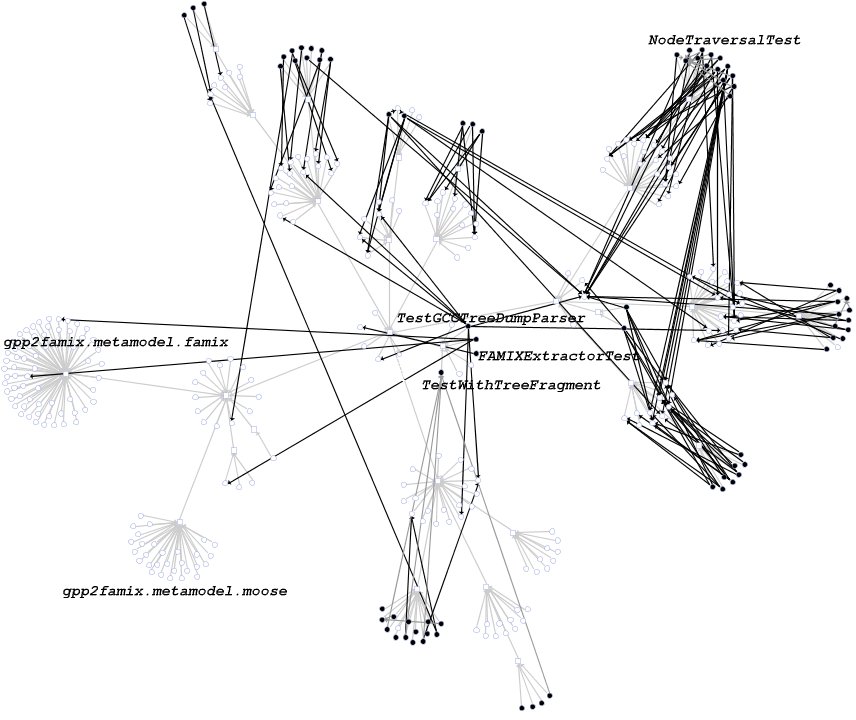}
	\caption{cpp2famix System-wide Test View}
\label{fig:cpp2famixsw}
\end{figure*}

\head{First Contact.} From \figref{fig:cpp2famixsw}, we derive a consistent, per production class unit testing approach. The unit test code resides in a subpackage \emph{test} of each component, except for four components that are weakly or even completely uncovered. Furthermore, we identify \emph{cpp2famix.test.TestGCCTreeDumpParser} as an integration test, exercising the parsing and filtering of a GCC tree dump into the system's internal tree representation. For test dependencies, we noticed \testhelpers \emph{cpp2famix.node.traversal.test.NodeTraversalTest} and \emph{cpp2famix.test.TestWithTreeFragment}, helping test cases with traversing AST representations and with composing small test data trees respectively. 

\head{Understand Units.} The developer points out six production classes that are currently being modified: \emph{ClassExtractor}, \emph{FieldExtractor}, \emph{FieldsIterator}, \emph{Attribute}, \emph{Clazz} and \emph{StatementIterator}. Using the Unit under Test Views, we identify and navigate to the test cases involved:
\begin{compactitem}
	\item \emph{Attribute} and \emph{Clazz} are not directly tested as they belong to generated code, conform to our earlier findings using the System-Wide view.
	\item \emph{ClassExtractor}, \emph{FieldIterator} and \emph{StatementIterator} are exercised by corresponding \emph{*Test} test cases.
	\item \emph{FieldExtractor} is covered by \emph{FieldsIteratorTest} as well as by \emph{FieldExtractorTest}.
\end{compactitem}

\begin{figure}[htbp]
	\centering
	\includegraphics[scale=0.2]{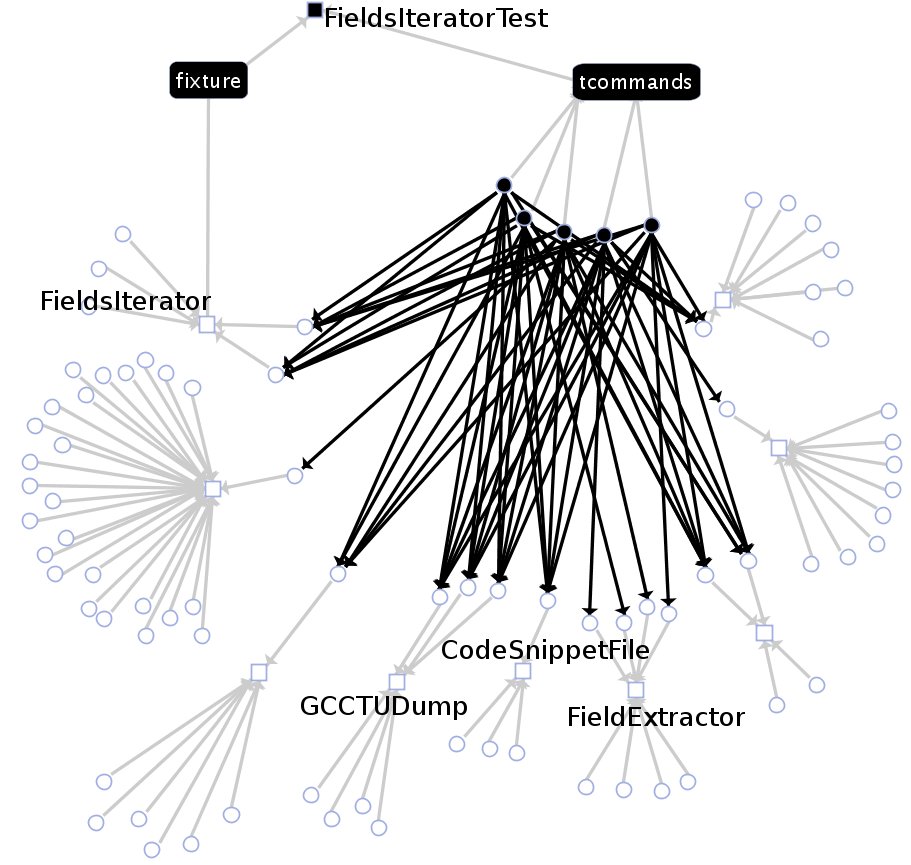}
	\vspace{-4mm}
	\caption{Test case FieldsIteratorTest}
	\vspace{-4mm}
\label{fig:cpp2famixfieldsiteratortest}
\end{figure}

\head{Assess Test Cases.} From the Test Case View, we deduce that quite some helper objects are needed to test the behavior of the Extractor and Iterator classes. The class names of the test helpers, however, reveals that sample pieces of data are composed to exercise the units under test (\figref{fig:cpp2famixfieldsiteratortest}). As such we conclude that these are broadly isolated unit tests.

\subsubsection{Validation}

During an interview, we confronted the system's developer with our analysis. He testifies that a tight unit testing approach (using \junit) has been undertaken, with test cases being written either just before (test-driven) or just after the corresponding production code. This results in a class and method coverage of 90\% and 79\% respectively. The developer acknowledges the presence of untested components, explaining that he did not see the need to test the generated components \emph{cpp2famix.metamodel.famix} and \emph{cpp2famix.metamodel.moose}. For the classes in \emph{cpp2famix.metamodel}, he commented that we were looking at dead code that was replaced by the classes in the two subpackages. \emph{cpp2famix.extractors.*}, at last, only contains simple data holders and as such it was not considered worthwhile to be unit tested.

\emph{cpp2famix.test.TestGCCTreeDumpParser} is confirmed to be an integration test, but the developer stated that it is an old test that even failed when we tried to execute it. Instead, he points to \emph{cpp2famix.test.FAMIXExtractorTest} as being the current integration test, although it is not implemented in a traditional \SSVT style, but merely a "user" of the top level production class. That explains why we did not identify this test case as one that deserves special attention.

\section{Related Work}
\label{sec:relatedWork}

We identified the following work in the domain of test suite reverse engineering.

Agrawal et al. introduce a set of techniques to enhance program understanding, debugging and testing \cite{621029}. Among others, the $\chi$Suds tool suite contains tools to assist developers in achieving high test coverage, locating errors as well as minimizing regression sets. Via source code coloring, the developer perceives the coverage level, erroneous locations or execution frequency.

Gaelli et al. observe that not all unit tests are alike \cite{Gael05a}. Therefore, a taxonomy that distinguishes unit tests based on the focus on one or more methods, type of expected outcome, etc. Their automated classification approach for SUnit tests using heuristics achieves a high overall precision (89\%) and a moderate recall (52\%). One of the steps the authors identify as future work involves making explicit the relationship between unit tests and methods under test.

Van Geet and Zaidman hypothesize that unit tests covering multiple units are less suited as documentation as such tests are harder to understand \cite{VanGeetPCODA2006}. In a case study involving the \ant project, the median number of methods executed by a test command is more than 200, which make them conclude that the test suite of this particular project is not well suited for documentation purposes.

To gain knowledge about the inner working of a software system, Cornelissen et al. use sequence diagrams obtained from test execution \cite{Cornelissen2005}. The use of abstraction, separation of test stages and stack depth limitations make such diagrams scalable.

\section{Conclusion \& Future work}
\label{sec:conclusion}

In this work, we proposed a visualization technique assisting re-engineers to explore the composition of an object-oriented system's unit test suite. We propose three graph-based views, representing (aspects of) a test suite in terms of the \SSVT cycle's test concepts. These views assist a re-engineer in building initial understanding and assessing opportunities and weaknesses for further evolution. We describe how certain visual indicators in these views reveal information about the location of test cases, the coverage level (in an exploration context) as well as the followed unit test strategy.

We validated the technique by means of two case studies. In the first one, we compared the results of our analysis with \ant's system documentation as well as with finding of other authors. Our initial findings regarding coverage, key system classes as well as test design were confirmed. Secondly, we investigated a system which has been developed with a tight unit testing approach. The lead developer of this system, \cppfamix, confirmed most of our claims about the test suite.

Based on these two case studies, we conclude that the visual exploration technique, as a first contact technique, serves its purpose. As a next step in the reverse engineering of test suites, we identify a need for finer-grained analysis, such as (i) obtaining actual coverage measurements via test execution and (ii) incorporating information about size and complexity of components for a more detailed assessment. We identify the integration of such information, e.g., via polymetric views, as future work. 

\vspace*{-1ex}
{\small
\paragraph{Acknowledgements --}
This work was executed in the context of the ITEA project if04032 entitled \emph{Software Evolution, Refactoring, Improvement of Operational\&Usable Systems} (SERIOUS) and has been sponsored by IWT, Flanders. 
}
\vspace*{-2ex}
\bibliographystyle{latex8}
\bibliography{biblio}
\end{document}